# High-intensity wave vortices around subwavelength holes: from ocean tides to nanooptics

Kateryna Domina[1], Pablo Alonso-González[2,3], Andrei Bylinkin[1,4], María Barra-Burillo[4], Ana I. F. Tresguerres-Mata[2,3], Francisco Javier Alfaro-Mozaz[4], Saül Vélez[4,5], Fèlix Casanova[4,6], Luis E. Hueso[4,6], Rainer Hillenbrand[6,7], Konstantin Y. Bliokh[1,6,8,*], Alexey Y. Nikitin[1,6,*]

[1]*Donostia International Physics Center (DIPC), Donostia-San Sebastián 20018, Spain*
[2]*Department of Physics, University of Oviedo, Oviedo 33006, Spain*
[3]*Center of Research on Nanomaterials and Nanotechnology, CINN (CSIC-Universidad de Oviedo), El Entrego 33940, Spain*
[4]*CIC nanoGUNE BRTA, Donostia - San Sebastian, 20018, Spain*
[5]*Departamento de Física de la Materia Condensada, Condensed Matter Physics Center (IFIMAC) & Instituto Nicolás Cabrera, Universidad Autónoma de Madrid, Madrid 28049, Spain*
[6]*IKERBASQUE, Basque Foundation for Science, Bilbao 48009, Spain*
[7]*CIC nanoGUNE BRTA and EHU/UPV, Donostia-San Sebastián, 20018, Spain*
[8]*Centre of Excellence ENSEMBLE3 Sp. z o.o., 01-919 Warsaw, Poland*
*alexey@dipc.org, konstantin.bliokh@dipc.org

Vortices are ubiquitous in nature; they appear in a variety of phenomena ranging from galaxy formation in astrophysics to topological defects in quantum fluids. In particular, *wave vortices* have attracted enormous attention and found applications in optics, acoustics, electron microscopy, etc. Such vortices carry quantized phase singularities accompanied by *zero intensity* in the center, and quantum-like orbital angular momentum, with the minimum localization scale of the *wavelength*. Here we describe a conceptually novel type of wave vortices, which can appear around arbitrarily small 'holes' (i.e., excluded areas or defects) in a homogeneous 2D plane. Such vortices are characterized by *high intensity* and confinement at the edges of the hole and hence *subwavelength* localization of the angular momentum. We demonstrate the appearance of such vortices in: (i) optical near fields around metallic nanodiscs on a dielectric substrate, (ii) phonon-polariton fields around nanoholes in a polaritonic slab, and (iii) ocean tidal waves around islands of New Zealand and Madagascar. We also propose a simple toy model of the generation of such subwavelength vortices via the interference of a point-dipole source and a plane wave, where the vortex sign is controlled by the mutual phase between these waves. Our findings open avenues for subwavelength vortex/angular-momentum-based applications in various wave fields.

## Introduction

Wave vortices [1] have been observed and employed in a variety of classical wave fields: optical [2,3], acoustic [4,5], plasmonic [6,7], elastic [8], water-surface [9], as well as in quantum matter waves: superfluidic [10,11], electron [12,13], neutron [14], atomic [15]. In particular, optical vortices have found applications in manipulation of small particles [16,17], super-resolution microscopy [18,19], quantum entanglement [20,21], information transfer [22,23], and astronomy/astrophysics [24,25]. Such universality and broad spectrum of applications can be explained by the combination of robust *topological* and effective *dynamical* properties of wave vortices, supported by the facility of their generation.

Here we argue that there can be two types of wave vortices with dramatically different mathematical and physical properties, which will be described below. While most of previous works deal with only one of these types, the second type has not been systematically studied and recognized (although vortices of the second type can be found in some earlier experimental works). To make this point clear, we start with a mathematical introduction of these two types of wave vortices.

First, conventional (type-I) wave vortices appear around *zeros* or *phase singularities* of a complex wavefield, i.e., points of vanishing intensity and phase winding by $2\pi\ell$ around them, where the integer number $\ell$ is called topological charge of the vortex [2,3]. Assuming the wavefield over a 2D plane, $\psi(x,y)$, and a circularly-symmetric intensity distribution with the zero point at the coordinate origin, the vortex has the form $\psi \propto f(r)e^{i\ell\varphi}$, where $(r,\varphi)$ are the polar coordinates and the radial function behaves as $f(r) \propto r^{|\ell|}$ near the origin, Fig. 1(a). Such wavefield is an eigenmode of the quantum-like orbital angular momentum (OAM) operator $\hat{L}_z = -i\partial/\partial\varphi$ with the eigenvalue $\ell$ [1,2]. Importantly, the zero intensity in the center and wave nature determine the *minimum radius* of the localization of energy in a wave vortex: about half of the *wavelength* for $|\ell| = 1$, Fig. 1(a).

The above-mentioned properties are well known for the usual 'dark-core' 2D wave vortices relying on the assumption of (i) a *finite-amplitude* wavefield over (ii) a *simply-connected* (i.e., without excluded areas) homogeneous plane. Remarkably, if we abandon one of these two conditions, another type of a wave vortex emerges. Indeed, a complex function can have phase winding not only around *zeros* but also around its *poles*, i.e., points of diverging amplitude. For example, the 2D Helmholtz equation $\nabla^2\psi + k^2\psi = 0$, describing a wavefield with a fixed frequency and wavenumber $k$, allows two types of circularly-symmetric solutions:

$$\psi_I \propto J_{|\ell|}(kr)e^{i\ell\varphi}, \qquad \psi_{II} \propto Y_{|\ell|}(kr)e^{i\ell\varphi}, \tag{1}$$

where $J_{|\ell|}$ and $Y_{|\ell|}$ are the Bessel functions of the first and second kind, which have a zero and a pole in the $r = 0$ center, respectively. The first solution (1) is the type-I vortex mode, described above and shown in Fig. 1(a). The second solution is typically neglected in problems with a finite wavefield. However, if the origin $r = 0$ is excluded from the consideration, i.e., we consider a not-simply-connected plane with a 'hole' in the center: $\mathbb{R}^2 \setminus \{r \leq a\}$, where $a$ is the radius of the hole, the second-type solution (1) becomes a finite-amplitude solution of the problem with suitable boundary conditions, Fig. 1(b). In what follows, we refer to such kind of solutions as *type-II vortices*.

The type-II vortex in (1) has the topological $2\pi\ell$ phase winding around the hole and the same OAM properties determined by the same azimuthal phase factor as in the type-I vortex. However, it also has remarkable *localization* properties. Namely, assuming a small subwavelength hole with $a \ll \lambda = 2\pi/k$ and the radial wavefunction behaving as $|\psi| \propto 1/r^{|\ell|}$, the energy in such vortex is concentrated near the hole boundary, i.e., confined to the *subwavelength* radius $\sim a$, which can be *arbitrarily small*, independently of the wavelength. We emphasize that type-I and type-II vortices are *topologically* different objects on a not-simply-connected 2D plane: while type-I vortices are associated with phase singularities, type-II vortices have well defined phases winding around finite-size holes in the plane.

The goal of this work is to demonstrate the appearance of type-II wave vortices localized around subwavelength holes in various physical systems. The role of such mathematical holes can be played by either a real hole in a surface supporting waves, or a nanoparticle/defect on a homogeneous surface, or an island in the sea with water-surface waves. We will show that 'bright' (high-intensity) vortices around subwavelength exclusions are generic objects, which can be just as important as the type-I dark-core vortices around intensity zeros. We will also describe the basic mechanisms of their generation. We believe that subwavelength localization of the vortex energy and OAM can find a broad range of applications from optical nanoscales to colossal geophysical scales.

## Dipole interference models

We now consider simple physical models involving point-dipole fields for the generation of subwavelength high-intensity (type-II) vortices. We first note that the wave field of a point dipole source decays as $1/r$ and thus diverges at the $r = 0$ origin. Therefore, to deal with physical fields of finite amplitude, we have to exclude some area around this singularity, i.e., introduce a 'hole'. Next, by interfering two dipolar sources located at the origin and oscillating in the orthogonal directions in the $(x,y)$ plane with a phase difference of $\pm\pi/2$, we have the field $\psi \propto (x \pm$ $iy)/r^2 = (1/r)e^{\pm i\varphi}$, which is exactly the type-II vortex with $\ell = \pm 1$ shown in Fig. 1(b). For instance, such interfering dipoles can be excited by illuminating a small circular defect (e.g., a hole) on a metal surface supporting surface plasmon-polaritons (SPP) by a circularly-polarized normally $z$-incident light [26,27]. The circular polarization interacting with the defect induces the corresponding circular dipole in the $(x, y)$ plane generating an SPP vortex outside the defect. Note, however, that this mechanism is essentially based on optical spin-orbit interactions [28], i.e., conversion of the spin (circular polarization) of light coming from the third dimension into the OAM of the SPP vortex. This involves 3D vector properties of electromagnetic waves rather than simple 2D scalar waves. Moreover, most plasmonic experiments [26,27] deal with defects of the SPP wavelength size and truly subwavelength confined vortices still remain elusive.

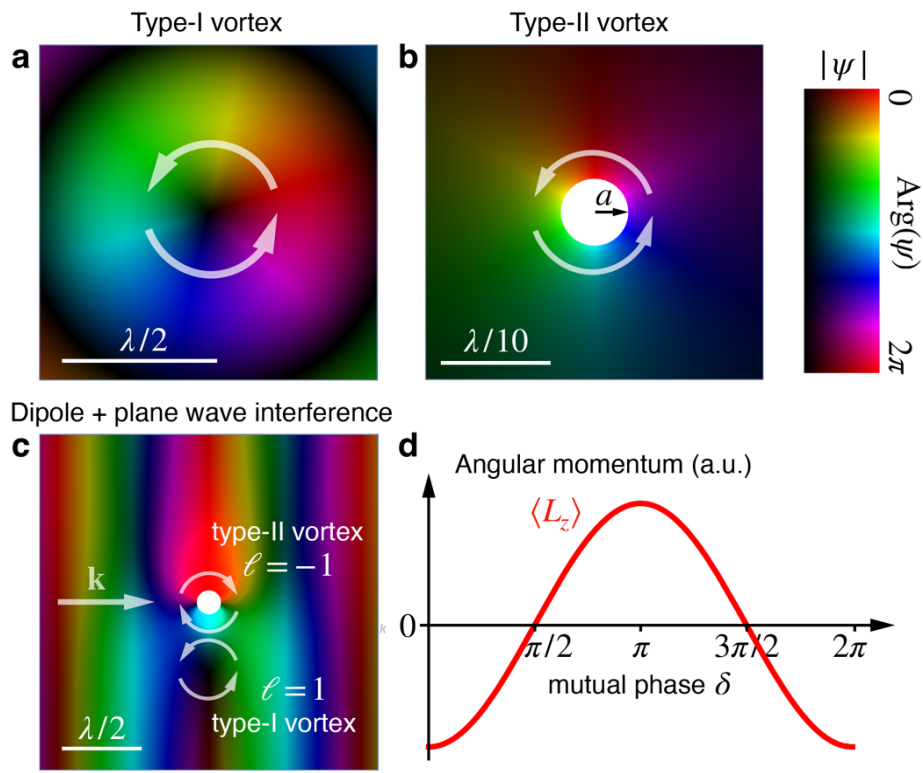

**Fig. 1. Usual type-I dark-core wave vortices vs. type-II high-intensity subwavelength vortices around holes.** **(a)** Circularly symmetric type-I vortex around the intensity zero (phase singularity) given by $\psi_I$ in Eq. (1) with $\ell = 1$. Here and hereafter the brightness and color in the plot correspond to the amplitude and phase of the wave field, as shown in the colorbar. The diameter of the intensity maximum is about the wavelength $\lambda$. The circular arrow shows the phase gradient (wave current) which generates the OAM $L_z$. **(b)** Circularly symmetric type-II vortex around a subwavelength circular hole (exclusion in the plane) of the radius $a = \lambda/20$, given by $\psi_{II}$ in Eq. (1) with $\ell = 1$. The diameter of the intensity maximum is the hole diameter $2a \ll \lambda$. Such type-II vortex can be considered as a residue of a pole of the wavefunction on the plane without exclusion. **(c)** Interference of a point-dipole source and a plane wave, Eq. (2), can produce a pair the type-I and type-II vortices of opposite topological charges $\ell = \pm 1$. Since the energy is mostly concentrated in the bright type-II vortex around the dipole, the total OAM of this field is nonzero. The OAM and vortices can be controlled by the mutual phase $\delta$ between the dipole and plane wave, as shown in **(d)**. The wave field in **(c)** corresponds to $\delta = 0$.

Here we put forth another model that also involves a dipole source, but does not require a circularly polarized excitation and, therefore, can be more relevant for 2D scalar wave systems. Namely, we consider an interference of a $y$-oriented point dipole and a plane wave propagating in the $x$ direction. The resulting interference field can be written as

$$\psi \propto A e^{ikx+i\delta} + \frac{y}{kr^2}, \qquad (2)$$

where $A$ and $\delta$ are the relative (with respect to the dipole) amplitude and phase of the plane wave. (Note that the dipole field in Eq. (2) can be regarded as an asymptotic form of the exact Hankel-function solution of the Helmholtz equation: $\psi \propto \frac{y}{r} H_1^{(1)}(kr) = \frac{y}{r}[J_1(kr) + iY_1(kr)]$ at $kr \ll 1$.) The phase-amplitude distribution of the wavefield (2) is shown in Fig. 1(c). Remarkably, this field exhibits both type-I and type-II vortices. For $\delta = 0$, the usual 'dark' type-I vortex with $\ell = 1$ appears around the zero-intensity point $(x, y) = (0, -(kA)^{-1})$, whereas the 'bright' (high-intensity) type-II vortex with $\ell = -1$ appears around the excluded origin area. Thus, the co-existence and difference between the two types of vortices is clearly seen in this model. The positions and existence of the two vortices can be controlled by the relative phase $\delta$. These vortices disappear for $\delta = \pi/2$ and $3\pi/2$ and flip their topological charges, $\ell \to -\ell$, for $\delta = \pi$ (the 'dark' vortex moves to $y = (kA)^{-1}$ in this case), see Supplementary Figure 1.

Far from the origin, the dipole field decays and the total wavefield (2) approximates to the plane wave. It is topologically trivial at this global scale, meaning that the phase increment along a closed contour embracing both of the vortices vanishes: the total topological charge of this wavefield is zero. Nonetheless, the field (2) carries nonzero OAM, because the contribution from the 'bright' type-II vortex prevails over the contribution from the 'dark' type-I vortex. To show this, note that the field (2) is not an OAM eigenmode, and one has to calculate the expectation (mean) value of the OAM: $\langle L_z \rangle = \langle \psi | \hat{L}_z | \psi \rangle \propto \iint \text{Im}(\psi^* \partial_\varphi \psi) dx dy$. Although this integral cannot be properly normalized (the total plane-wave energy diverges), its value is nonzero and behaves as $\propto -\cos \delta$, see Fig. 1(d). This evidences that the total OAM is largely determined by the contribution from the subwavelength type-II vortex near the origin. Notice the difference between the zero total topological charge and the nonzero total OAM in our system: these are generally independent properties of the wave field, coinciding only for the circularly-symmetric OAM eigenmodes [29].

The above models demonstrate that subwavelength high-intensity type-II vortices can emerge naturally in 2D wave systems with point sources requiring exclusion of the divergence points. Below we provide examples of such vortices in real wave systems of strikingly different nature.

## Tidal-wave vortices around islands

Probably the earliest (in the 1830s) known example of the observation of phase singularities or type-I wave vortices is provided by lunar semi-diurnal (M2) ocean tidal waves [30], which are described by a 2D scalar wave field over the globe. Phase singularities with zero tidal amplitude, known as *amphidromic points* [31], occur naturally throughout the world ocean, Fig. 2(a). Importantly, studying modern M2 tidal maps [32], we found the high-intensity type-II vortices around big islands (which play the role of 'holes' in the ocean): New Zealand and Madagascar, see Fig. 2(c) and Supplementary Figure 2. Although these islands have dimensions of $a \sim 1{,}000$ km, these are actually *subwavelength* exclusions: the wavelength of M2 tides can be estimated, assuming shallow-water gravity waves with the period of $T = 12.4$ hours and ocean depth of $h = 5$ km, as $\lambda \simeq 10{,}000$ km. Fig. 2(c) clearly shows that the tidal-wave energy is concentrated around these islands on a subwavelength scale in the form of type-II vortices with $\ell = 1$.

There is a vast oceanographic literature on coastal (e.g. Kelvin) waves around islands [33,34]. In particular, Longuet-Higgins, considering shallow-water gravity waves around round islands [33], theoretically pointed out the possibility of high-intensity vortex solutions $\psi_{II}$, Eq. (1), when the Coriolis force is negligible. (When the Coriolis force is large, the corresponding solution involves the modified radial Bessel function $K_{|\ell|}$ with similar properties near the island.) Thus, the tidal vortices around New Zealand and Madagascar provide examples of subwavelength high-intensity type-II wave vortices.

Notably, our dipole and plane wave interference model (2) with adjusted parameters provides a very good approximation to the tidal wave maps around these islands, as shown in Fig. 2 and Supplementary Figure 2. A possible explanation of this agreement is that these massive islands oscillate under the Moon gravity force with the same M2 frequency (see the Earth tides [35]) and hence act as dipolar sources, which interfere with tidal near-plane wave propagating in the ocean around the island.

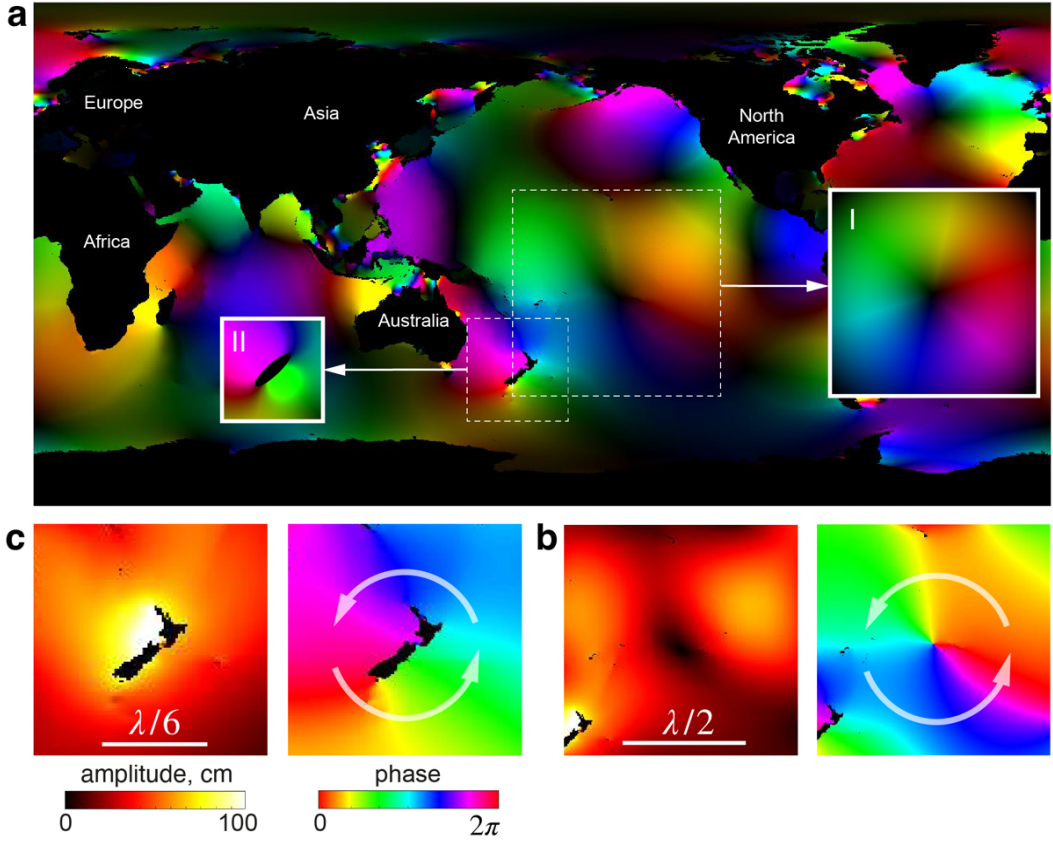

**Fig. 2. Low-intensity type-I vortices around amphidromic-points and high-intensity type-II vortices around islands in M2 tidal waves. (a)** Amplitude-phase map of the M2 tidal waves from the HAMTIDE model data [32]. **(b)** Zoomed-in amplitude and phase distributions for the type-I vortex around the amphidromic point in the middle of the Pacific Ocean. **(c)** Zoomed-in intensity and phase distributions for the type-II vortex around the New Zealand islands. Insets I and II in panel **(a)** show the model type-I and type-II vortices from Fig. 1(a) and (c) with suitably adjusted parameters.

## Nanophotonic experiments

We now demonstrate the appearance of subwavelength type-II vortices in nanooptical systems, such as subwavelength disc scatterers and holes in a 2D layer supporting *phonon-polaritons (PhP)*, i.e., hybridized excitations of light and lattice vibrations.

In the first scenario, we employ the excitation of optical near fields around subwavelength metal discs to demonstrate the two dipole-interference models described above. In the simulation shown in Fig. 3(a), we illuminate a gold disc (radius $a = 0.75$ μm, on a CaF2 substrate) by a normally-incident circularly-polarized electromagnetic plane wave (the wavelength $\lambda = 7$ μm). The electric field of this wave has the form of $E_x \bar{\mathbf{x}} \pm i E_y \bar{\mathbf{y}}$ (the overbar denotes the unit vectors), and it excites two orthogonal $\pi/2$-phase-shifted dipoles in the $(x, y)$ plane in the disc. The interference of these dipoles results in a near field with the $z$-component bearing the corresponding vortex, $E_z \propto e^{\pm i\varphi}$, and quickly decaying away from the disc. This transformation of circular polarization into a vortex can be attributed to the spin-orbit interaction of light [28]. Figure 3(a) shows the amplitude and phase distributions of the near field $E_z(x,y)$ in the plane above the disc, which clearly exhibits a subwavelength-confined type-II vortex similar to Fig. 1(b). In contrast to the diverging mathematical point dipole, the disc has a finite size naturally removing the amplitude singularity. In our approach, the disc can be regarded as an exclusion area ('hole') in free space, so that the field around it represents the type-II vortex solution (1) for the free-space wave equation. Remarkably, the vortex arises independently of the size of the scatterer (see Supplementary Figure 3), and thus it can be confined to an arbitrarily small scale. The spin-orbit mechanism of the vortex generation essentially involves 3D vector properties of light and spin angular momentum (circular polarization) of the incoming light [6,7,26–28]. For a linear polarization of the incoming light, $E_y = 0$, a usual single dipole is excited, without a vortex or OAM, as shown in Fig. 3(b).

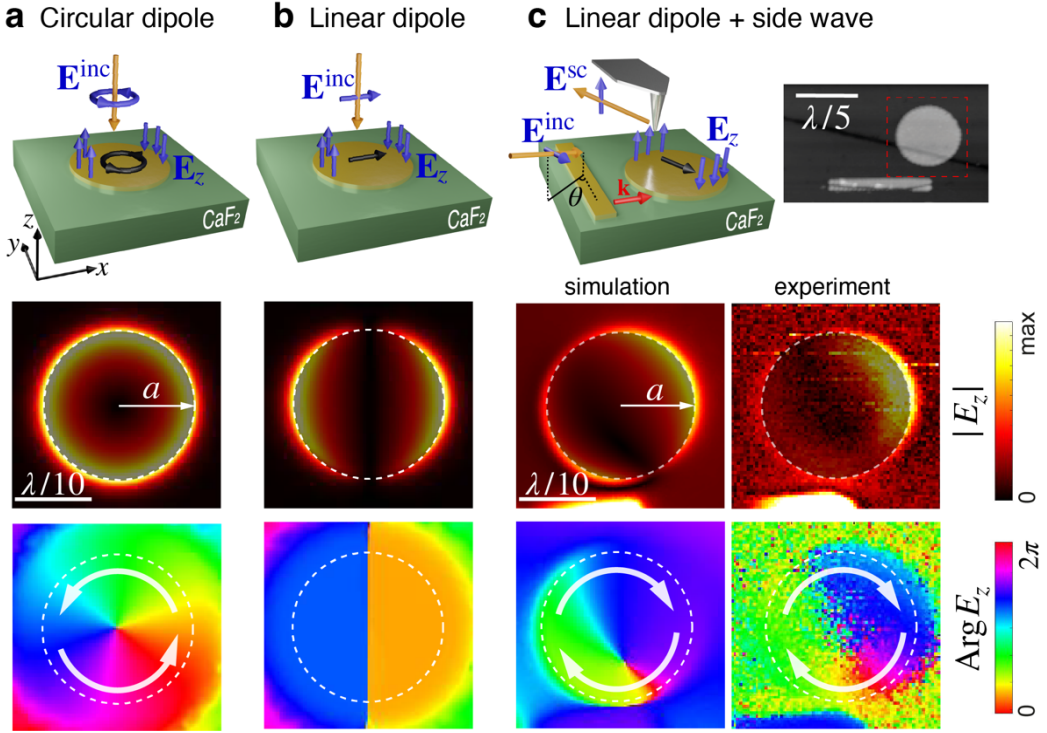

**Fig. 3. Generation of subwavelength type-II optical vortices around metallic discs via excitation of a circular dipole (3D spin-orbit interaction) and via 2D interference of a linear dipole with a side wave.** The top panels in **(a-c)** provide schematics of the systems considered. The black arrows on the discs indicate the circular and linear dipoles excited by the incident light. The lower panels show the results of numerical simulations and experimental measurements [in **(c)**] for the amplitude and phase distributions of the excited electric near field component $E_z(x,y)$ above the disc. **(a)** A subwavelength gold disc on a dielectric substrate illuminated by normally incident circularly-polarized light. The excited electric near field $E_z(x,y)$ above the disc exhibits a type-II vortex (considering the disc area as an exclusion in free space). Its formation is explained by the spin-orbit interaction in the 3D vector optical field [26–28]. **(b)** Same as **(a)** but for linearly-polarized light does not produce a vortex. **(c)** Same as **(b)** but with a rod antenna placed next to the gold disc and obliquely incident light (the inset in the top panel shows the sample topography obtained by AFM). A type-II vortex is formed due to the interference of the wave propagating from the antenna and the linear dipole induced at the disc, similarly to the scalar 2D model in Eq. (2). Simulations (left column) exhibit a nearly perfect agreement with experiment (right column).

Next, we perform an experiment demonstrating the second, spin-independent way to generate a subwavelength type-II vortex. For this purpose, we fabricated a gold disc with the same radius $a = 0.75$ μm on a CaF$_2$ substrate and placed a 2.3 μm long gold rod next to it, see Fig. 3(c) and Methods for details. When illuminated by an obliquely incident light (linearly s-polarized in our experiment), the gold rod acts as a resonant antenna launching a wave propagating orthogonally to it, while the gold disc produces a dipolar electric field. In the $(x, y)$ plane, this effectively mimics the interference of a dipole and a propagating side wave, modelled by Eq. (2). We imaged the near-field distribution in the $(x, y)$ plane using scattering-type scanning near-field microscopy (s-SNOM, see Methods) [36]. Notably, as s-SNOM probe we use a Si dielectric tip [Fig. 3(c)], allowing visualization of the near-field distribution without substantial field distortion (typical s-SNOM experiments are carried out using metal tips that both excite and probe the near fields). While scanning the sample, both the amplitude and phase of the $E_z(x, y)$ component of the p-polarized tip-scattered light is recorded simultaneously with the sample topography.

As a result, we observed a clear vortex above the disc with the intensity localized near its edge. This vortex is similar to the shown in Fig. 3(a) but, importantly, its generation did not require the spin angular momentum of the incoming light. Equation (2) and Fig. 1(c,d) evidence that such vortices and nonzero OAM can appear within purely scalar 2D wave fields. Note that due to the weak signal from the Si tip, we had to adjust the mutual positions of the disc and antenna, as well as the angle of incidence of the incoming light. For the same reason, we made sure that the disc size was large enough (but still in the subwavelength regime).

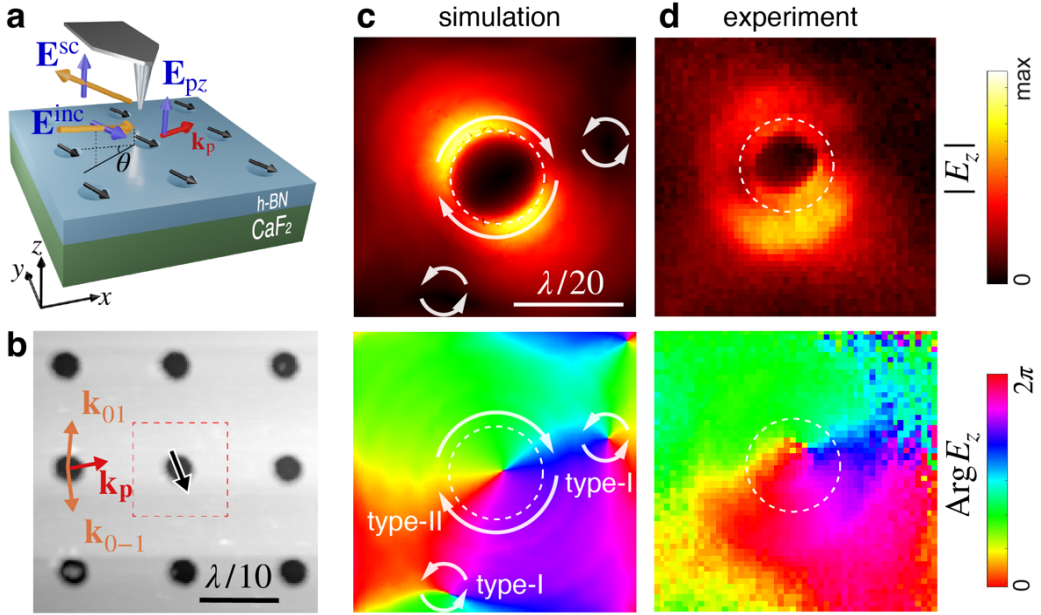

**Fig. 4. Generation of subwavelength type-II phonon-polariton vortices around nanoholes. (a)** Schematics of the experiment involving a PhP crystal on a h-BN layer excited by a linearly s-polarized obliquely-incident light and measured by s-SNOM. The phonon-polariton wavevector $k_p$ and excited dipoles at the holes are shown by the red and black arrows, respectively. **(b)** AFM topography of the PhP crystal. **(c)** Numerical simulations of the near field amplitude (top) and phase (bottom) of the near field component $E_z(x, y)$ of a Bragg PhP mode around a single hole of the PhP crystal. **(d)** Experimental s-SNOM measurements of the near-field distributions around a single hole of the PhP crystal. In both, simulations and experiment, one can clearly see a high-intensity type-II vortex confined around the subwavelength hole.

In the second scenario, we find subwavelength type-II vortices around nanoholes in a thin van der Waals crystal layer (h-BN) supporting PhPs [37,38]. (Note that the usual type-I vortices were observed in PhPs only very recently [39,40].) An h-BN layer supports a set of hyperbolic PhP modes, M$n$, with the integer $n$ characterizing the number of the field oscillations across the layer [41]. Although the h-BN layer has a strong $z$-oriented uniaxial anisotropy, the hyperbolic PhP modes are isotropic in the $(x,y)$ plane. In a thin h-BN layer (a few orders of magnitude thinner than the wavelength), the fundamental mode M0 dominates, and PhPs are strongly $z$-confined $(x,y)$-propagating waves.

We first perform numerical simulations of PhPs in a structure involving a single nanohole and a metal nanoantenna next to it, similar to the configuration in Fig. 3(c). The resulting plot (Supplementary Figure 4) exhibits a clear type-II vortex around the nanohole. However, the experimental realization of this configuration is challenging due to the weak dipole field induced at a single hole. Therefore, we performed experimental measurements in a structure with periodically-arranged multiple holes – a polaritonic crystal (PC) [42], Fig. 4(a,b). The dipole excitations of the individual holes can be significantly enhanced due to collective effects, thereby dramatically favoring the signal-to-noise ratio in the near-field measurements. Namely, at certain frequencies suchPC exhibits Bragg resonances, in which polaritons form Bloch modes of high intensity (see Supplementary Figure 5) carrying in-plane momentum acquired from the obliquely incident light. The interference between the propagating Bloch wave and the dipoles induced in the holes can produce type-II vortices, according to the model (2).

The PC is formed by a square array of circular holes (radius $a = 150$ nm), with a period of 900 nm, designed to exhibit PhP Bragg resonances in the mid-IR range (wavelength 6.76–7.35 μm). The holes are etched in an h-BN layer (thickness $h = 38$ nm) on a transparent $CaF_2$ substrate (see Methods). The Bloch mode (0,1), originating from the fundamental M0 PhP waveguiding mode, was excited by obliquely incident light of wavelength $\lambda = 7$ μm, elevation angle of incidence $\gamma = 30°$, and azimuthal angle $\theta = 22°$. As in previous experiments, we used s-SNOM to visualize the electric field of the excited PhP.

Figures 4(c,d) show the intensity and phase near-field images in the vicinity of a single hole obtained experimentally and in numerical simulations. In both cases, one can clearly see a subwavelength type-II vortex of topological charge $\ell = -1$ confined at the hole edge. Note that the intensity distribution contains a signature of the dipole field induced at the hole. We have numerically calculated the normalized expectation value of the OAM of the field $E_z(x,y)$ in the unit cell: $\langle L_z \rangle = \iint \text{Im}(E_z^* \partial_\varphi E_z) dx dy / \iint |E_z|^2 dx dy$. This resulted in $\langle L_z \rangle \simeq -0.342$ for the simulations and $\langle L_z \rangle \simeq -0.591$ for the experiment, which is consistent with the imperfect high-intensity vortex of topological charge $\ell = -1$. Note also that the simulations in Fig. 4(b) show phase singularities, i.e., type-I vortices, at some distance from the hole, similar to Fig. 1(c). Such type-I vortices are ubiquitous for low-intensity field zones, and they do not contribute much to the total OAM of the field, determined by the high-intensity type-II vortex.

## Concluding remarks

We have described theoretically and observed experimentally a new type of wave vortices which appear around 'holes' (i.e., exclusions or defects) in a 2D plane. The usual well-studied (type-I) wave vortices in a homogeneous space are formed around intensity zeros (phase singularities) with their energy distributed on the wavelength or larger scales. In sharp contrast to this, the type-II vortices are characterized by high intensity confined around arbitrarily small (subwavelength) exclusions. Mathematically, the two types of vortices can be associated with zeros and poles (truncated by the exclusion) of a complex wave function.

Notably, we have shown that type-II vortices can naturally appear in strikingly different systems: from colossal ocean tidal waves to tiny phonon-polariton waves. The Supplementary Video shows temporal dynamics of the instantaneous real fields given by $\text{Re}[\psi(\mathbf{r})e^{-i\omega t}]$ for a complex wave field $\psi(\mathbf{r})$ with frequency $\omega$, in the physical systems considered in this work: (i) the dipole + wave interference model (2), (ii) M2 tides around New Zealand, and (iii) phonon-polariton waves around a nanohole. All these examples exhibit entirely similar vortex-induced rotations of subwavelength-confined fields.

We have described the main models for the generation of type-II vortices via interference of two orthogonal phase-shifted dipole sources (a circular dipole) or a single linear dipole and a side-coming plane wave. Here the dipole source is associated with the 'hole' in the plane, and the vortex is formed around it. In different systems, the role of such hole/dipole can be played by different subwavelength inhomogeneities: an island in water surface, a metallic nanoparticle in a dielectric layer, or a hole in a thin layer supporting surface plasmon/polariton waves. In all these cases, one can clearly see the general features of type-II vortices and their distinction from the usual type-I vortices.

Importantly, when the type-I and type-II vortices co-exist in the same wave field [e.g., Fig. 1(c), Fig. 2, and Fig. 4(c)], they appear as 'dark' and 'bright' vortices, respectively. Therefore, the energy and OAM is largely concentrated in type-II vortices. Such vortices provide subwavelength confinement of the wave OAM, which can be highly important for various applications, such as vortex microlasers [43,44], high-harmonic generation with OAM [45], and vortex generation at nanoscales [46]. In this work we have only provided the basic concept, main properties, and examples of scalar type-II vortices in various wave systems. There can be numerous extensions of this problem for future studies: vector properties of type-II vortices, their appearance in laboratory fluidic and acoustic systems, the case of multiple 'holes' located close to each other, etc. Given the enormous importance of type-I vortices in a variety of classical and quantum waves, we believe that the type-II vortices open the avenue for new subwavelength vortex-based phenomena in different fields.

**Acknowledgements:** The authors are greatly thankful to Prof. Gabriel Molina-Terriza for the discussions. The authors acknowledge support from the Spanish Ministry of Science and Innovation (grants PID2019-111156GB-I00, PID2020-115221GB-C42, PID2021-122980OA-C53, PID2021-122511OB-I00, PID2021-123949OB-I00, PID2022-141304NB-I00 and PRE2020-092758 funded by MCIN/AEI/10.13039/501100011033 and by 'ERDF—A Way of Making Europe'). A.I.F.T.-M. acknowledges support from the Severo Ochoa program of the Government of the Principality of Asturias (no. PA-21-PF-BP20-117). P.A.-G. acknowledges support from the European Research Council under Consolidator grant No. 101044461, TWISTOPTICS. A.Y.N. acknowledges the Basque Department of Education (grant PIBA-2023-1-0007) and a 2022 Leonardo Grant for Researchers in Physics, BBVA Foundation. The Foundation takes no responsibility for the opinions, statements and contents of this project, which are entirely the responsibility of its authors. K.Y.B. and R.H. acknowledge the International Research Agendas Programme (IRAP) of the Foundation for Polish Science co-financed by the European Union under the European Regional Development Fund and Teaming Horizon 2020 program of the European Commission [ENSEMBLE3 Project No. MAB/2020/14]; the TEAM program of the Foundation for Polish Science co-financed by the European Union under the European Regional Development Fund [Grant No. TEAM/2016-3/29]. R.H., F.C., L.E.H., A.B., M.B.-B., F.J.A.-M., S.V. acknowledges support from the Spanish Ministry of Science and Innovation under the Maria de Maeztu Units of Excellence Program (CEX2020-001038-M/MCIN/AEI/10.13039/501100011033).

**Author contributions:** A.Y.N. originally conceived the study, basing on the data from preliminary nanoimaging experiments by F.J.A.-M. and P.A.-G., and suggested the toy model. K.Y.B. put forward the general concept of type-II vortices around subwavelength 'holes' and examples with tidal waves. K.D. performed the simulations and modelling. K.D., P.A.-G., A.B, A.I.F.T.-M. and F.J.A.-M. performed the s-SNOM nanoimaging measurements. F.J.A.-M. performed the far-field measurements of the polaritonic crystal. M.B.B. and S.V. fabricated the samples under the supervision of F.C. and L.E.H. K.D., R.H., K.Y.B. and A.Y.N. discussed vortex concepts and interpreted the results. K.Y.B. and A.Y.N. supervised the project and co-wrote the manuscript with the input from all the co-authors.

# References

1. J. F. Nye and M. V. Berry, "Dislocations in wave trains," *Proc. R. Soc. A* **336** 165 (1974).
2. L. Allen, S. M. Barnett, and M. J. Padgett (Eds.), *Optical Angular Momentum* (IoP Publishing, Bristol, 2003).
3. J. P. Torres and L. Torner (Eds.), *Twisted Photons* (Wiley-VCH, Weinheim, 2011).
4. B. T. Hefner and P. L. Marston, "An acoustical helicoidal wave transducer with applications for the alignment of ultrasonic and underwater systems," *J. Acoust. Soc. Am.* **106**, 3313 (1999).
5. S. Guo, Z. Ya, P. Wu, and M. Wan, "A review on acoustic vortices: Generation, characterization, applications and perspectives," *J. Appl. Phys.* **132**, 210701 (2022).
6. Y. Gorodetski, A. Niv, V. Kleiner, and E. Hasman, "Observation of the Spin-Based Plasmonic Effect in Nanoscale Structures," *Phys. Rev. Lett.* **101**, 043903 (2008).
7. E. Prinz, M. Hartelt, G. Spektor, M. Orenstein, and M. Aeschlimann, "Orbital Angular Momentum in Nanoplasmonic Vortices," *ACS Photonics* **10**, 340 (2023).
8. G. J. Chaplain, J. M. De Ponti, and T. A. Starkey, "Elastic orbital angular momentum transfer from an elastic pipe to a fluid," *Commun. Phys.* **5**, 279 (2022).
9. N. Francois, H. Xia, H. Punzmann, P. W. Fontana, and M. Shats, "Wave-based liquid-interface metamaterials," *Nat. Commun.* **8**, 14325 (2016).
10. E. J. Yarmchuk, M. J. V. Gordon, and R. E. Packard, "Observation of stationary vortex arrays in rotating superfluid helium," *Phys. Rev. Lett.* **43**, 214 (1979).
11. A. L. Fetter, "Rotating trapped Bose-Einstein condensates," *Rev. Mod. Phys.* **81**, 647 (2009).
12. J. Verbeeck, H. Tian, and P. Schattschneider, "Production and application of electron vortex beams," *Nature* **467**, 301 (2010).
13. K. Y. Bliokh, I. P. Ivanov, G. Guzzinati, L. Clark, R. Van Boxem, A. Beche, R. Juchtmans, M. A. Alonso, P. Schattschneider, F. Nori, and J. Verbeeck, "Theory and applications of free-electron vortex states," *Phys. Rep.* **690**, 1 (2017).
14. C. W. Clark, R. Barankov, M. G. Huber, M. Arif, D. G. Cory, and D. A. Pushin, "Controlling neutron orbital angular momentum," *Nature* **525**, 504 (2015).
15. A. Luski, Y. Segev, R. David, O. Bitton, H. Nadler, A. R. Barnea, A. Gorlach, O. Cheshnovsky, I. Kaminer, and E. Narevicius, "Vortex beams of atoms and molecules," *Science* **373**, 1105 (2021).
16. H. He, M. E. J. Friese, N. R. Heckenberg, and H. Rubinsztein-Dunlop, "Direct observation of transfer of angular momentum to absorptive particles from a laser beam with a phase singularity," *Phys. Rev. Lett.* **75**, 826 (1995).
17. D. G. Grier, "A revolution in optical manipulation," *Nature* **424**, 810 (2003).
18. F. Tamburini, G. Anzolin, G. Umbriaco, A. Bianchini, and C. Barbieri, "Overcoming the Rayleigh criterion limit with optical vortices," *Phys. Rev. Lett.* **97**, 163903 (2006).
19. M. Ritsch-Marte, "Orbital angular momentum light in microscopy," *Phil. Trans. R. Soc. A* **375**, 20150437 (2017).
20. A. Mair, A. Vaziri, G. Weihs, and A. Zeilinger, "Entanglement of orbital angular momentum states of photons," *Nature* **412** 313 (2001).
21. J. Leach, B. Jack, J. Romero, A. K. Jha, A. M. Yao, S. Franke-Arnold, D. G. Ireland, R. W Boyd, S. M. Barnett, and M. J. Padgett, "Quantum correlations in optical angle–orbital angular momentum variables," *Science* **329**, 662 (2010).
22. N. Bozinovic, Y. Yue, Y. Ren, M. Tur, P. Kristensen, H. Huang, A. E. Willner, and S. Ramachandran, "Terabit-Scale Orbital Angular Momentum Mode Division Multiplexing in Fibers," *Science* **340**, 1545 (2013).
23. M. Krenn, J. Handsteiner, M. Fink, R. Fickler, R. Ursin, M. Malik, and A. Zeilinger, "Twisted light transmission over 143 km," *Proc. Natl. Acad. Sci. USA* **113**, 13648 (2016).
24. J. H. Lee, G. Foo, E. G. Johnson, G. A. Swartzlander, "Experimental verification of an optical vortex coronagraph," *Phys. Rev. Lett.* **97**, 053901 (2006).

25. F. Tamburini, B. Thidé, G. Molina-Terriza, and G. Anzolin, "Twisting of light around rotating black holes," *Nat. Phys.* **7**, 195 (2011).
26. Y. Gorodetski, S. Nechayev, V. Kleiner, and E. Hasman, "Plasmonic Aharonov-Bohm effect: Optical spin as the magnetic flux parameter," *Phys. Rev. B.* **82**, 125433 (2010).
27. G. M. Vanacore, G. Berruto, I. Madan, E. Pomarico, P. Biagioni, R. J. Lamb, D. McGrouther, O. Reinhardt, I. Kaminer, B. Barwick, H. Larocque, V. Grillo, E. Karimi, F. J. García de Abajo, and F. Carbone, "Ultrafast generation and control of an electron vortex beam via chiral plasmonic near fields," *Nat. Mater.* **18**, 573 (2019).
28. K. Y. Bliokh, F. J. Rodríguez-Fortuño, F. Nori, and A. V. Zayats, "Spin-orbit interactions of light," *Nat. Photonics* **9**, 796 (2015).
29. M. V. Berry and W. Liu, "No general relation between phase vortices and orbital angular momentum," *J. Phys. A: Math. Theor.* **55**, 374001 (2022).
30. M. V. Berry, "Making waves in physics," *Nature* **403**, 21 (2000).
31. D. E. Cartwright, *Tides: A Scientific History* (Cambridge University Press, 2000).
32. E. Taguchi, D. Stammer, and W. Zahel, "Inferring deep ocean tidal energy dissipation from the global high-resolution data-assimilative HAMTIDE model," *J. Geophys. Res. Oceans* **119**, 4573 (2014).
33. M. S. Longuet-Higgins, "On the trapping of long-period waves round islands," *J. Fluid Mech.* **37**, 773 (1969).
34. K. H. Brink, "Coastal-trapped waves and wind-driven currents over the continental shelf," *Annu. Rev. Fluid Mech.* **23**, 389 (1991).
35. P. Melchior, "Earth tides," *Surv. Geophys.* **1**, 275 (1974).
36. T. Neuman, P. Alonso-González, A. Garcia-Etxarri, M. Schnell, R. Hillenbrand, and J. Aizpurua, "Mapping the near fields of plasmonic nanoantennas by scattering-type scanning near-field optical microscopy", *Laser & Phton. Rev.* **9**, 637 (2015).
37. D. N. Basov, M. M. Fogler, and F. J. García de Abajo, "Polaritons in van der Waals materials," *Science* **354**, 195 (2016).
38. E. Galiffi, G. Carini, X. Ni, G. Álvarez-Pérez, S. Yves, E. M. Renzi, R. Nolen, S. Wasserroth, M. Wolf, P. Alonso-Gonzalez, A. Paarmann, and A. Alù, "Extreme light confinement and control in low-symmetry phonon-polaritonic crystals," *Nat. Rev. Mater.* **9**, 9 (2024).
39. M. Wang, G. Hu, S. Chand, M. Cotrufo, Y. Abate, K. Watanabe, T. Taniguchi, G. Grosso, C.-W. Qiu, and A. Alù, "Spin-orbit-locked hyperbolic polariton vortices carrying reconfigurable topological charges," *eLight* **2**, 12 (2022).
40. Y. Kurman, R. Dahan, H.H. Shenfux, G. Rosolen, E. Janzen, J.H. Edgar, F.H.L. Koppens, and I. Kaminer, "Dynamics of optical vortices in van der Waals materials," *Optica* **10**, 612 (2023).
41. S. Dai, Z. Fei, Q. Ma, A.S. Rodin, M. Wagner, A.S. McLeod, M.K. Liu, W. Gannett, W. Regan, K. Watanabe, T. Taniguchi, M. Thiemens, G. Dominguez, A.H. Castro Neto, A. Zettl, F. Keilmann, P. Jarillo-Herrero, M.M. Fogler, D.N. Basov, "Tunable Phonon Polaritons in Atomically Thin van der Waals Crystals of Boron Nitride," *Science* **343**, 1125 (2014).
42. F. J. Alfaro-Mozaz, S. G. Rodrigo, P. Alonso-González, S. Vélez, I. Dolado, F. Casanova, L. E. Hueso, L. Martín-Moreno, R. Hillenbrand, and A. Y. Nikitin, "Deeply subwavelength phonon-polaritonic crystal made of a van der Waals material," *Nat. Commun.* **10**, 42 (2019).
43. P. Miao, Z. Zhang, J. Sun, W. Walasik, S. Longhi, N. M. Litchinitser, and L. Feng, "Orbital angular momentum microlaser," *Science* **353**, 464 (2016).
44. C. Huang, C. Zhang, S. Xiao, Y. Wang, Y. Fan, Y. Liu, N. Zhang, G. Qu, H. Ji, J. Han, L. Ge, Y. Kivshar, and Q. Song, "Ultrafast control of vortex microlasers," *Science* **367**, 1018 (2020).
45. G. Gariepy, J. Leach, K. T. Kim, T. J. Hammond, E. Frumker, R. W. Boyd, and P. B. Corkum, "Creating High-Harmonic Beams with Controlled Orbital Angular Momentum," *Phys. Rev. Lett.* **113**, 153901 (2014).
46. Q. Chen, G. Qu, J. Yin, Y. Wang, Z. Ji, W. Yang, Y. Wang, Z. Yin, Q. Song, Y. Kivshar, and S. Xiao, "Highly efficient vortex generation at the nanoscale," *Nat. Nanotechnol.* (2024). https://doi.org/10.1038/s41565-024-01636-y